\documentstyle[12pt]{article}
\begin{document}

\vspace{10mm}
\hspace{20mm}
Published in: Mod. Phys. Lett. A, 1997, Vol. 12, No. 32, pp. 2475-2479

\begin{center}
{{\bf ENERGY SPECTRUM OF ANYON IN THE COULOMB FIELD        \\
\vspace{5mm}
             S.I. Kruglov and M.N. Sergeenko }              \\
\vspace{2mm}
{\it Academy of Sciences of Belarus, Institute of Physics   \\
                   Minsk 220072, Belarus } }
\end{center}    \vspace{5mm}

\begin{abstract}
Anyonic atom is considered as a two-dimensional system. Using some
approximations we find the energy spectrum of the anyon in the
Coulomb field. It is shown that the anyonic atom is stable.  \\ ~ \\
\noindent PACS number(s): 03.65.Ge, 03.65.Sq 
\end{abstract}

One of the interesting fundamental phenomenon which was observed in
the last decades is the discovery of anyons, relativistic spinning
particles in $2+1$ dimensions. In contrast to three-dimensional
space, indistinguishable quantum particles in two-dimensional space
can, in general, have anomalous statistics [1-4]. These
quasiparticles carry not only a charge $q$ also the magnetic flux
$\Phi_0$.

During the last several years the ($2+1$)-dimensional physics has
been an area of intense activity, mainly due to the application of
anyons in realistic planar physics [5]. This has provided the impetus
for constructing viable models of free anyons in analogy to the Dirac
equation for spin $1/2$ particles [6-8]. However, many interesting
properties of anyons, in particular, their gyromagnetic ratio $g$,
can only be probed in the presence of electromagnetic interactions
[9,10].

There is a problem to construct the relativistic wave equations for
anyons interacting with the electromagnetic field in $2+1$
dimensions. A new simple wave equation for an anyon was proposed in
Ref. [9]. To derive this equation the simplectic framework for the induced
representation of the Poincar\'e group for anyons, minimally coupled
to an external electromagnetic field, has been used.

In the present paper, we have investigate the Coulomb interaction of
anyons. Unlike Ref. [9], we define the spin operator to be $S_\mu =
-Sp_\mu /m$, where $S$ is the fractional spin and $m$ is the rest
mass of the anyon. Such redefinition has some advantages and allows to
obtain an approximate analytic expression for the bound state
spectrum of the two-body Coulomb problem in the presence of an
anyonic vector potential.

We start with the following equation of motion for the anyon in $2+1$
dimension:

\begin{equation}
\left( D_\mu ^2-\frac{2qSi}mF_\mu D_\mu -m^2\right) \phi  = 0,
\end{equation}
where $q$ is the charge of the anyon, $D_\mu  = \partial _\mu
-iqA_\mu $, $F_\mu  = \frac 12\epsilon _{\mu \nu \alpha }F_{\nu
\alpha }$, $\epsilon _{\mu \nu \alpha }$ is the antisymmetric tensor
($\epsilon _{012} = 1$), $F_{\mu \nu } = \partial _\mu A_\nu
-\partial _\nu A_\mu $ is the strength tensor, $A_\mu $ is the
vector-potential of the external electromagnetic field.  The equation
of motion (1) is the relativistic and gauge invariant equation like
the Klein-Gordon equation.  But here there is an additional second
term which gives the spin-orbit interaction. The gyromagnetic ratio
of the anyon in this scheme is $g = 2$ [9-11].

In two dimensional space, the vector potential $A_\mu $ to be $ A_k =
0$ ($k = 1,\,2$), $A_0 = -Ze/r$, where $r = (x_1^2+x_2^2)^{1/2}$. We
assume that the hydrogen-like atom consists of the anyon of charge
$q$ and mass $m$ and a nucleus of charge $Ze$ and large mass $M\gg
m$. Then Eq. (1) takes the form:

\begin{equation} \left[ \partial _k\partial ^k-\partial
_t^2+\frac{2i\xi Z}r\partial _t+\frac{ \xi ^2Z^2}{r^2}-\frac{4\xi
ZS}{mr^3}[i(x_2\partial _1-x_1\partial _2) ]-m^2\right] \psi (\vec r)
= 0, \end{equation}
where $\xi  = eq$ is the coupling constant.

By introducing the polar coordinates $x_1 = r\cos \varphi $, $x_2 =
r\sin \varphi $ and replacing the angular momentum operator $\hat J =
-i(x_2\partial _1-x_1\partial _2)$ by $\hat J = i\frac \partial
{\partial \varphi }$ we may take solution having the form

\begin{equation} \psi (r,\varphi ) = e^{-iEt}\sum_{-\infty }^\infty
e^{il\varphi }f_l(r), \end{equation}
where $l$ is the orbital quantum number and $E$ is the energy of the
anyon. Substituting (3) into Eq. (2) we obtain

\begin{equation} \left[ \frac 1r\frac d{dr}\left(
r\frac d{dr}\right) + E^2-m^2+\frac{2\xi EZ}r-\frac{l^2-\xi
^2Z^2}{r^2} - \frac{4\xi ZSl}{mr^3}\right] f_l(r) = 0.
\end{equation}
Units are chosen such that $\hbar = c = 1$.

Thus the problem under consideration has reduced to the radial
Schr\"odinger type equation (4) which determines the energy
eigenvalues of the anyonic atom. However, Eq. (4) represents the
so-called "insolvable" problem. By "solvable" models we mean those
models for which the eigenvalue problem can be reduced to a
hypergeometric function by a suitable transformation. Equation (4) is
not reduced to the hypergeometric equation.  Thus a different
approach is necessary if we are to find solution of Eq. (4).

One of the earliest and simplest methods of obtaining the approximate
eigenvalues of the one-dimensional problems is the WKB method. Proofs
of varying degrees of rigor have been advanced which demonstrate the
exactness of the WKB quantization condition \cite{Lang}-\cite{Se}.
It is well known that the standard WKB approximation in leading order
in $\hbar $ {\em always} reproduces exact energy spectrum for the
"solvable" spherically symmetrical potentials, $V(r)$, if the Langer
correction $L^2\rightarrow L^2+\frac 14$, $ L^2 = l(l+1)$ \cite{Lang}
($l$ is the orbital quantum number) in the centrifugal term of the
radial Schr\"odinger equation has fulfilled. One can easily to check
that analogous Langer type correction of the form $l^2\rightarrow
l^2+\frac 14$ in the leading order WKB quantization condition results
in the exact energy eigenvalues for the nonrelativistic two-dimension
Coulomb problem \cite{Ho}:  $E_n = -\frac 12\alpha ^2Z^2m(n'+l+\frac
12)^{-2}$.

Within the framework of the WKB method the solvable potentials mean
those potentials for which the eigenvalue problem has two turning
points. The radial problem (4) has three turning points that
represent a serious difficulty. To estimate the energy eigenvalues of
the anyonic atom let us consider Eq. (4) as a quasi-solvable problem
in the framework of the WKB method.

For this by writing $f_l = U_l(r)/\sqrt r$ we reduce Eq. (4) to the
canonical form. Then making the Langer-type replacement,
$l^2\rightarrow l^2+ \frac 14$, we obtain the second-order equation
in the required form:

\begin{equation} \left[ \frac{d^2}{dr^2} + E^2-m^2+\frac{2\xi
ZE}r-\frac{\lambda ^2} {r^2}-\frac{ 4\xi ZSl'}{mr^3}\right] U_l(r) =
0, \end{equation}
where $\lambda ^2 = l^2-\xi ^2Z^2$, $l' = (l^2+\frac 14)^{1/2}$, $
U_l(r) = f_l(r)\sqrt r$. Let us define the radial quantum number,
$n^{\prime } $, to be a number of zeros of the radial wave function
in the physical region, i.e., at $r>0$. Turning points of the
problem, $r_1$, $r_2$, $r_3$, ($r_1<r_2<r_3$) are defined as roots of
the cubic equation

\begin{equation} (E^2-m^2)r^3 + 2\xi ZEr^2 - \lambda ^2r - 4\xi
SZl^{\prime }m^{-1} = 0.  \end{equation}
Analysis of Eq.  (6) shows that only two turning points, $r_2$,
$r_3$, lye in the physical region, i.e., at $r>0$, and the turning
point $r_1$ lyes in the nonphysical region $r<0$.  Therefore one can
consider the leading order WKB quantization condition to be
appropriate at the interval $[r_2,\,r_3]$:

\begin{equation} \int_{r_2}^{r_3}\sqrt{E^2-m^2 + \frac{2\xi ZE}r -
\frac{\lambda ^2}{r^2}-\frac{4\xi SZl'}{mr^3}}dr  = \pi (n' + \frac 12).
\end{equation}

To calculate integral (7) let us note that the term $\sim r^{-3}$ in
this integral is essential at very small $r$, i.e., its contribution
can be considered as a small perturbation at the interval
$[r_2,r_3]$.  This means that the phase-space integral (7) can be
represented at this interval in the form:

\begin{equation}
\int _{r_2}^{r_3}\sqrt{k^2+\frac{2\xi ZE}r-\frac{\lambda ^2}{r^2}}dr
- \frac{2\xi SZl'}m\int _{r_2}^{r_3}\frac 1{\sqrt{k^2r^2 + 2\xi
ZEr-\lambda ^2}}\frac{dr}{r^2}  = \pi (n' + \frac 12), \end{equation}
where the turning points $r_2,$ $r_3$ are determined approximately
from the equation $k^2r^2+2\xi ZEr-\lambda ^2 = 0$, $k^2 = E^2-m^2$.
Simple integration of (8) gives the equation for $E$:

\begin{equation} \frac{\xi ZE}{\sqrt{-E^2+m^2}} - \frac{2\xi
^2Z^2Sl^{\prime }} {\lambda ^3}\frac Em  =  n' + \frac 12 + \lambda .
\end{equation}

This equation results in the fourth degree equation with respect to
$E$.  However Eq. (9) can be simplified. For this note that the
second term in the left-hand side of the Eq. (9) arising from the
spin-orbit interaction is a small correction of the order $\sim \xi
^2$ at the interval under consideration. Therefore, if we represent
$E = m + E'$, where $E'$ is the kinetic energy which is of order
$\sim \xi ^2$ for the Coulomb potential, then this term can be
written approximately in the form $ \sigma _l = 2\xi ^2Z^2Sl^{\prime
}\lambda ^{-3}$. Now, the subsequent simple calculations result in
the energy eigenvalues for the anyonic atom:

\begin{equation} E_n  = m\left[ 1+\frac{\xi ^2Z^2}{(n' + \frac 12 +
\sqrt{l^2-\xi ^2Z^2} +\sigma _l)^2}\right]^{-\frac 12}.
\end{equation}

Let us analyze the result obtained. The formula (10) is analogous to
the relativistic energy eigenvalues for the Coulomb problem obtained
from the Klein-Gordon equation. The difference is in the structure of
the principal quantum number which contains the additional quantity
$\sigma _l$. If we write the principal quantum number in (10) with
the accuracy up to $\xi ^2$, $N\simeq n'+\frac 12+l +
[2Sl^{-2}(l^2+\frac 14)^{1/2}-\frac 12]\xi^2Z^2l^{-1}$, then one can see
that the additional term $\tilde \sigma_l = 2\xi ^2Z^2Sl^{-3}
(l^2+\frac 14)^{1/2}$ can be considered as the anyon fractional spin which
is added to the angular momentum $l$.  This term is the relativistic
correction; in nonrelativistic approximation, the formula (10)
results in the known exact result for the two-dimensional atom.

To check formula (10) we compare (see Table 1) the  eigenvalues
calculated for kinetic energy $E' = E - m$ with the corresponding
numerical calculations. As an example, in these calculations, we
have considered the case $S=1/2$, $\xi = \alpha$ and $m = m_e$ (electron
mass).

\begin{center}Table 1. {\em The energy spectrum  of anyonic atom}
\end{center}
\begin{center}
\begin{tabular}{llll} \hline
$n'$\hspace{5mm} & $l$ \hspace{10mm} & $E'_{anal}$,\ eV &
$E'_{num}$,\ eV  \\
\hline
$0$ & $1$ & $-6.0464$ & $-6.0467$  \\
$0$ & $2$ & $-2.1769$ & $-2.1769$  \\
$1$ & $1$ & $-2.1768$ & $-2.1770$  \\
$1$ & $2$ & $-1.1107$ & $-1.1106$  \\
$2$ & $1$ & $-1.1106$ & $-1.1106$  \\
$2$ & $2$ & $-0.6719$ & $-0.6718$  \\
\hline \end{tabular} \end{center}
We see that formula (10) reproduces the exact eigenvalues to
three-place accuracy.

To conclude let us summarize the obtained results. In this work, the
anyonic atom have considered as a two-dimensional system. Using the
semiclassical approximation we have found the energy eigenvalues of
the atom. There have shown that the anyonic atom is stable and has
the complex structure of the energy spectrum which is connected with
the fractional spin of the anyon.

{\em Acknowledgments}. This work was supported in part by the
Belarussian Fund for Fundamental Researches.

\end{document}